\documentclass[
reprint,amsmath,amssymb,aps,pra,floatfix,]{revtex4-1}
\usepackage{graphicx}	
\usepackage{dcolumn}	
\usepackage{bm}            
\usepackage{color}

\begin{document}
\title{
Ultrahigh-precision
measurement of the 
$n$$=$$2$ 
triplet 
P
fine structure of atomic helium using 
frequency-offset 
separated oscillatory fields
}
\author{K. Kato}
\author{T.D.G. Skinner}
\author{E.A. Hessels} 
\email{hessels@yorku.ca}
\affiliation{Department of Physics and Astronomy, York University, Toronto,
Ontario M3J 1P3, Canada}

\date{\today} 

\begin{abstract}

For decades, 
improved 
theory and experiment 
of the 
$n$$=$$2$ 
$^3$P 
fine structure of 
helium have allowed for 
increasingly-precise 
tests of 
quantum electrodynamics, 
determinations of the 
fine-structure 
constant
$\alpha$,
and limitations on 
possible
beyond-the-Standard-Model
physics.
Here we use the new 
frequency-offset
separated-oscillatory-fields
(FOSOF)
technique 
to measure the  
$2^3$P$_2\!\!\to\!2^3$P$_1$
interval. 
Our result of 
$2\,291\,176\,590(25)$~Hz
represents a major step forward in precision
for helium 
fine-structure
measurements.

\begin{description}
\item[PACS numbers]
\verb+\pacs{32.70.Jz,32.80.-t}+
\end{description}

\end{abstract}

\pacs{Valid PACS appear here}
\maketitle


In 1964, 
Schwartz 
suggested 
\cite{PR.134.A1181} 
that a 
part-per-million 
(ppm)
determination
of the 
fine-structure 
constant 
$\alpha$ 
might be possible using the 
2$^3$P 
fine structure of atomic helium if 
advances were made to both 
theory and experiment. 
In the past five decades, 
great progress has indeed been made 
in experimental measurements 
\cite{PR.169.55,
PRL.26.1613,
PRA.24.264,
PRA.24.279,
PRL.72.1802,
IEEETransInstrMeas.44.518,
PRL.84.4321,
PRL.105.123001,
OpticsComm.125.231,
PRL.82.1112,
PRL.92.023001,
PRL.97.139903,
CanJPhys.83.301,
PRL.95.203001,
PRA.58.R8,
PRL.84.3274,
PRL.87.173002,
PRA.79.060503,
PRA.91.030502,
zheng2017laser}
(including the evaluation 
of new systematic effects
\cite{PRA.86.012510,
PRA.86.040501,
PRA.89.043403,
marsman2015quantum,
marsman2015effect})
and in the 
quantum-electrodynamic 
(QED)
theory
\cite{PR.137.A1672,
PR.140.A1498,
PRA.5.2027,
PRA.6.865,PRA.7.479,
AnnPhysNY.82.89,
PRL.29.12,
PRA.18.867,
PRL.74.4791,
PRA.54.1252,
PRA.53.3896,
PRL.77.1715,
CanJPhys.80.1195,
JPhysB.33.5297,
JPhysB.36.803,
JPhysB.43.074015,
JPhysB.32.137,
PRL.97.013002,
PRA.79.062516,
PRA.80.019902,
PRL.104.070403,
CanJPhys.89.95,
JPhys.264.012007,
pei2015precision}
of these intervals.
The present measurement 
of the 
$2^3$P$_2\!\!\to\!2^3$P$_1$
interval has an uncertainty of only 
25~Hz,
which,
when compared to the full 
31.9~GHz 
2$^3$P
fine structure,
for the first time 
reaches
below 
one part per billion 
(ppb).
Thus, the present measurement is the first 
building block towards using the
2$^3$P 
fine structure for
tests of physics and fundamental constants 
at the 
$<$1-ppb 
level.
The other building blocks necessary 
will be a measurement of the the 
$2^3$P$_1\!\!\to\!2^3$P$_0$
interval
(which can be done with the same method
demonstrated in this work)
and an advance of
QED
theory to this same level of accuracy.
The latter will require 
extending 
the already heroic 
calculation of 
Pachucki 
and 
Yerokhin
\cite{PRL.104.070403}
to one order higher in 
$\alpha$. 

The payoff from assembling all of the 
building blocks will be large. 
First, 
the comparison between experiment and 
theory will provide the most accurate test
to date of 
QED
in a multielectron system
\cite{pachucki2017testing}.
Second, 
a
$<$1-ppb
test of the 
2$^3$P
fine structure will directly 
test 
(at 100 times the current accuracy)
for 
beyond-the-Standard-Model
physics
\cite{pachucki2017testing}, 
such as exotic 
spin-dependent 
interactions between electrons
\cite{ficek2017constraints}.
Third, 
the combination of 
$<$1-ppb
theory and experiment
would allow for a determination of the 
$\alpha$ 
at a level of 
$<$0.5~ppb, 
which is approaching the level of the current 
best determinations of 
$\alpha$
based on the electron magnetic moment
($g_e$)
\cite{hanneke2008new,
aoyama2012tenth,
laporta2017high,
aoyama2018revised}
and atomic recoil
\cite{parker2018measurement,
bouchendira2011new}.
Comparing values of 
$\alpha$
determined from various systems 
allows for tests of
beyond-the-Standard-Model
physics in each of the systems
\cite{bouchendira2011new,
hanneke2008new}.
In particular, 
the 
$g_e$
measurement,
given another determination of 
$\alpha$,
becomes a 
0.25-ppb 
test of 
QED,
and tests
for possible substructure of the electron
\cite{bouchendira2011new,
gabrielse2014precise}
and
the possible presence of dark photons
\cite{bouchendira2011new,
hanneke2008new,
kahn2017light},
and puts limits on possible 
dark axial vector bosons
\cite{bouchendira2011new,
hanneke2008new}.
The recoil measurement,
along with another 
$\alpha$
determination,
could be used for an
absolute mass standard 
\cite{lan2013clock}.
 
The current work is the first 
implementation of the new 
frequency-offset
separated-oscillatory-fields 
(FOSOF)
technique
\cite{vutha2015frequency},
which is a modification of the 
Ramsey 
method 
\cite{ramsey1949new}
of 
separated oscillatory fields 
(SOF).
For 
FOSOF,
the frequencies 
of the two separated fields 
are slightly offset from each other, 
so that the relative phase of the 
two fields varies continuously with time.

\begin{figure}[t!]
\includegraphics[width=3.5in]
{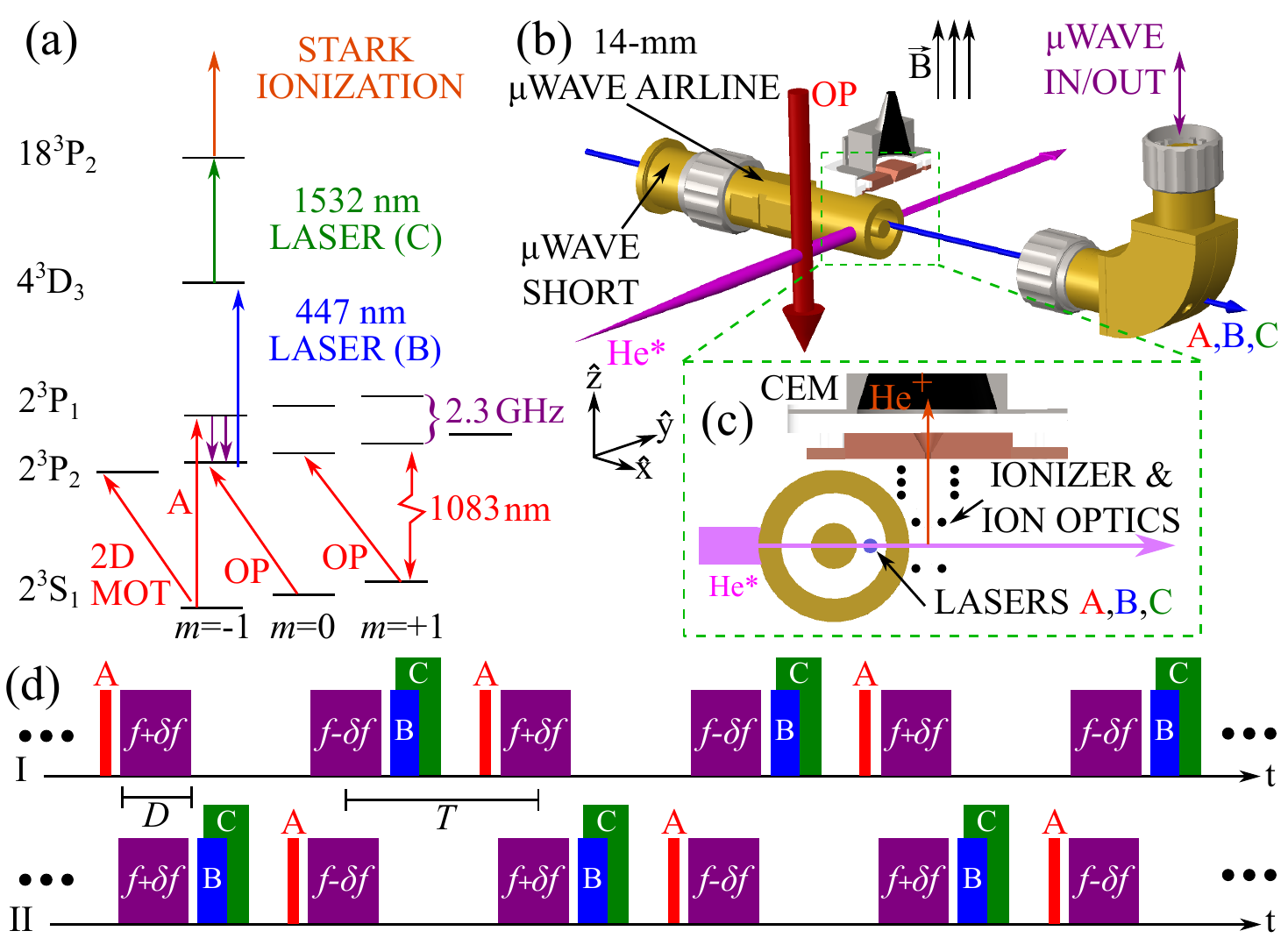}
\caption{\label{fig:ExptSetup} 
The experimental setup for the measurement.
An
energy-level
diagram 
(a) 
shows the 
2.3-GHz interval being measured
and the laser
transitions used.
The experimental setup 
(b), 
along with an expanded view of 
the region where the measurement
takes place 
(c)
shows the laser and microwave 
interactions and ionization detection.
The timing diagrams 
(d) 
show the 
FOSOF 
microwave pulses. 
}
\end{figure}

Our measurement uses a beam of metastable
2$^3$S
atoms created in a 
liquid-nitrogen-cooled
DC 
discharge source.
Two-dimensional
magneto-optical 
trapping
(2DMOT),
using permanent magnets and 
interactions with 
1083-nm
laser beams from the 
four transverse directions
(that interact for 
20 cm
along the 
atomic beam)
concentrates the 
beam to a flux of 
2$^3$S 
atoms of
7$\times$10$^{12}$/cm$^2$/s.
The atoms in this beam are 
optically pumped 
(OP 
in 
Fig.~\ref{fig:ExptSetup})
into the 
2$^3$S($m$=$-$1) 
state before passing through a 
0.5-mm-high 
slit into a coaxial microwave
airline.
The measurement takes place inside this airline.
The 
2$^3$S($m$=$-$1)
atoms are excited by a 
15-ns
pulse of 
1083-nm
laser light 
(A 
in 
Fig.~\ref{fig:ExptSetup})
up to the
2$^3$P$_1$($m$=$-$1)
state.
The 
$2^3$P$_1(m$=$-$1$)\!\!\to\!2^3$P$_2(m$=$-$1)
transition is then driven
with microwaves.
The resulting
2$^3$P$_2(m$=$-$1)
atoms are detected by exciting them up to
4$^3$D$_3$($m$=$-$1)
using a 
50-ns
pulse of
linearly-polarized
447-nm
laser light
(B
in 
Fig.~\ref{fig:ExptSetup}),
and then to
18$^3$P$_2$($m$=$-$1)
using a 
80-ns
pulse of 
linearly-polarized
1532-nm light
(C
in 
Fig.~\ref{fig:ExptSetup}).
The 
18$^3$P$_2$
atoms are 
Stark-ionized
by electric fields created using 
the wires shown in 
Fig.~\ref{fig:ExptSetup}(c),
and the resulting ions are focused 
through a 
1-mm
slit into a 
channel electron multiplier
(CEM).
The CEM current is dominated by 
ions created by these steps, 
with only a very small background 
from collisional ionization at 
our ultra-high-vacuum pressure of 
3$\times$10$^{-9}$~torr.

\begin{figure}[t!]
\includegraphics[width=3.5in]
{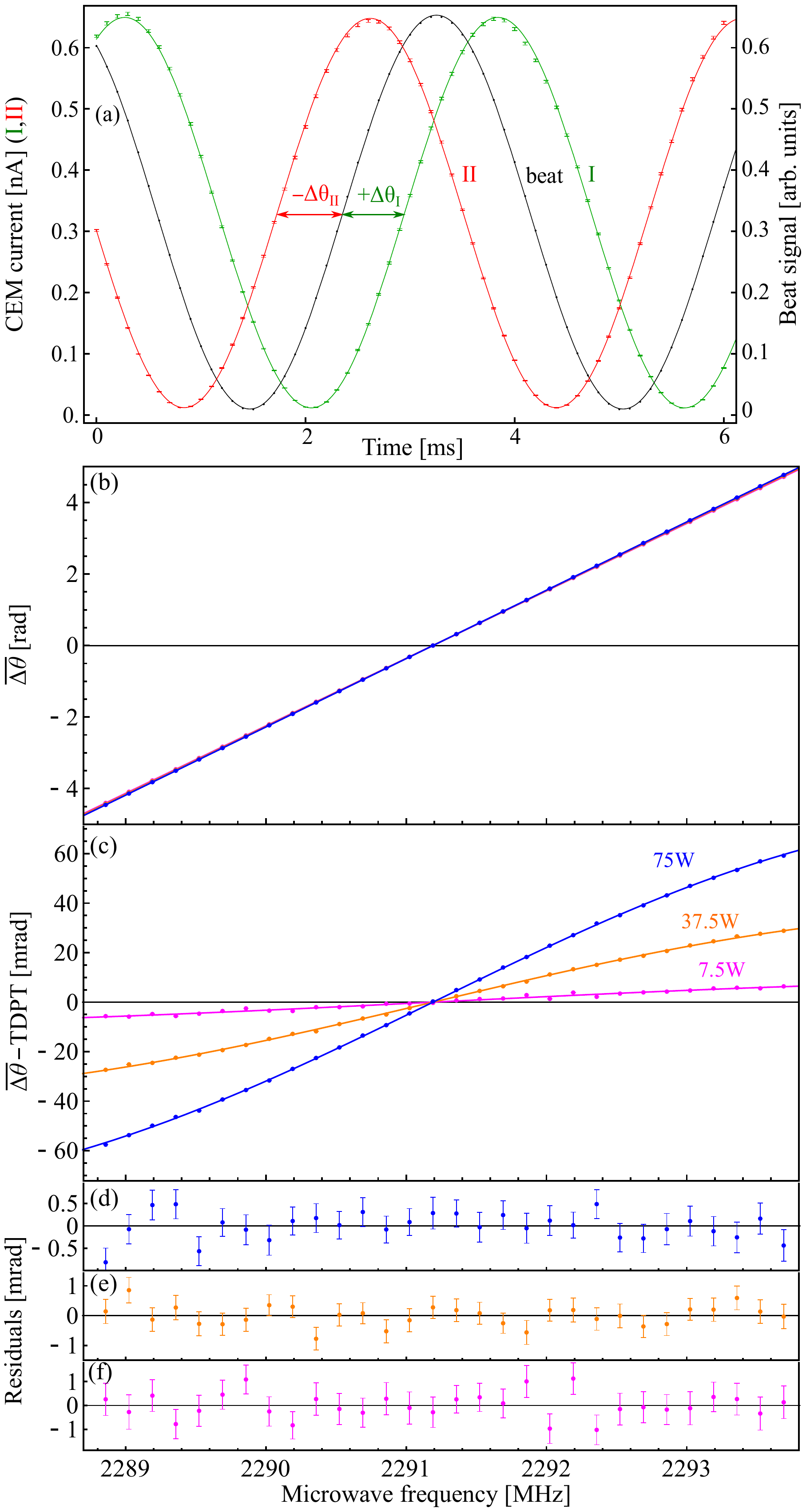}
\caption{\label{fig:LineShape} 
The FOSOF lineshape.
The sinusoidal atomic signals for 
case~I
and 
II
of 
Fig.~\ref{fig:ExptSetup}(d)
are shifted by 
$\Delta\theta_I$
and
$\Delta\theta_{II}$ 
relative to a 
microwave beat signal, 
as shown in 
(a).
The average phase shift 
$\overline{\Delta\theta}$ 
is shown in 
(b)
for powers of
7.5,
37.5,
and
75~W.
A 
100 
times expanded scale in 
(c),
where the 
straight-line
predicted by 
TDPT 
is subtracted,
resolves the lineshapes
for different powers.
The fits in 
(c) 
use
Eq.~(\ref{eq:LineShape}),
and the residuals from the fits 
are shown in 
(d), 
(e),
and 
(f).
}
\end{figure}

All three wavelengths
are produced using diode lasers.
The 
1083-nm
and 
1532-nm
light is amplified using fiber amplifiers.
The pulses are created using double passes 
through 
acousto-optic 
modulators.
The transition being measured 
is driven with two pulses of microwaves,
each of duration 
$D$, 
and separated in time by 
$T$,
as shown in 
Fig.~\ref{fig:ExptSetup}(d).
The pulses are created by fast switching of
outputs from two 
precision microwave generators,
with their internal clocks locked to 
each other and referenced to both 
Rb 
and 
GPS
clocks.
The microwaves enter one end of the airline,
and reflect off of a short 
to form a standing wave.
The returning wave is
monitored on a power detector and 
an oscilloscope.
The microwave frequencies
of the pulses are offset by
$\pm\delta\! f$,
with pulses alternating between 
$f$$+$$\delta\! f$
and 
$f$$-$$\delta\! f$.
The offset frequency 
$\delta\! f$
causes the relative phases of the two
pulses to vary continuously in time.
As a result,
the atomic signal 
(see 
Fig.~\ref{fig:LineShape}(a))
varies sinusoidally in time, 
cycling between destructive and 
constructive interference.
The phase difference 
$\Delta\theta$
between this 
signal and a beat signal
obtained by combining the microwaves
at the two frequencies 
is shown in 
Fig.~\ref{fig:LineShape}(a).
We take data with 
two different timing sequences
(Fig.~\ref{fig:ExptSetup}(d)):
with the 
$f$$+$$\delta\! f$ 
pulse before the 
$f$$+$$\delta\! f$ 
pulse 
(I),
and vice versa 
(II).
To switch from  
I
to
II,
only the timing of the laser pulses
is changed -- 
the microwave pulses are untouched.
Fig.~\ref{fig:LineShape}(a)
shows that the direction of the
phase shift 
$\Delta\theta$
is opposite for the two cases,
and,
as a result, 
the average  
$\overline{\Delta\theta}
$=($\Delta\theta_I$$-$$
\Delta\theta_{II}$)/2
cancels 
unintended phase shifts due to 
lags in either the atomic or beat
signals
\cite{vutha2015frequency}.
 
For the simple case of a two-level
system with two ideal pulses of 
duration 
$D$ 
and separation 
$T$,
the 
Schr$\ddot{\rm o}$dinger 
equation predicts 
a FOSOF lineshape 
$\overline{\Delta \theta} (f)$
of 
\begin{equation}
\label{eq:LineShape}
\Delta\omega(T\!-\!D)
\!+\!2\tan\!^{-1}\!
\big[
\frac
{\Delta \omega 
\tan(\sqrt{4 V^2\!+\!\Delta\omega^2}D/2)}
{\sqrt{4 V^2\!+\!\Delta\omega^2}}
\big],
\end{equation}
where 
$V$ 
is the
magnetic-dipole
matrix element driving the transition,
and 
$\Delta \omega$/2$\pi$$=$$f$$-$$f_0$ 
is the separation between the applied microwave
frequency and the atomic resonant frequency. 
This line shape is antisymmetric with respect to
$\Delta \omega$, 
and reduces to simply
$\Delta \omega T$
for small
$V$, 
as can also
be derived from 
time-dependent
perturbation theory 
(TDPT).
The observed lineshape is shown in 
Fig.~\ref{fig:LineShape}(b) for 
the case of 
$T$$=$300~ns
and
$D$$=$100~ns
for three different powers,
$P$.
The three powers give 
almost-identical,
almost-linear 
lineshapes.
On a
100-times 
expanded scale in 
Fig.~\ref{fig:LineShape}(c),
where the 
TDPT straight line has been 
subtracted,
one can see that the data is described 
well by the lineshape of 
Eq.~(\ref{eq:LineShape}).

The signal-to-noise ratio
(S/N)
for the 
measurement is astonishingly good, 
due to the large number of metastable 
atoms afforded by the 
2DMOT, 
due to the
near-unity 
efficiency for detection via 
Stark
ionization,
and due to the lack of a background
in our signal, 
which leads to uncertainties  
limited only by 
the shot noise in the signal itself.
The excellent 
S/N
can be seen directly in 
Fig.~\ref{fig:LineShape}(a), 
where each point is an average of 
20~ms of data.
Note that the uncertainties in these
plots are considerably larger at the top of the sinusoidal
signal than at the bottom, 
as the 
shot-noise 
limited uncertainties are proportional
to the square root of the signal size.
The resulting 
lineshapes of
Fig.~\ref{fig:LineShape}(b),
also show the excellent 
S/N, 
with each point representing 
40~s of data. 
A fit to the 
75-W 
data gives 
an
$f_0$
determination
with an uncertainty of only 
30 Hz.
This excellent 
S/N 
is despite the fact that measurement
sequence takes 
450~ns
(much longer than the
98-ns 
2$^3$P
lifetime),
allowing only 
$e^{-(450\, {\rm ns})/(98\, {\rm ns})}$=1.0\%
of the 
2$^3$P
atoms to contribute to the signal. 
The excellent 
S/N
allows for a
$T$ 
of up to
900~ns,
where only
22~ppm 
of the 
2$^3$P
atoms contribute.
When all of the data used for
this measurement is averaged,
the statistical uncertainty is 
$<$2~Hz.

\begin{figure*}[t!]
\includegraphics[width=7in]
{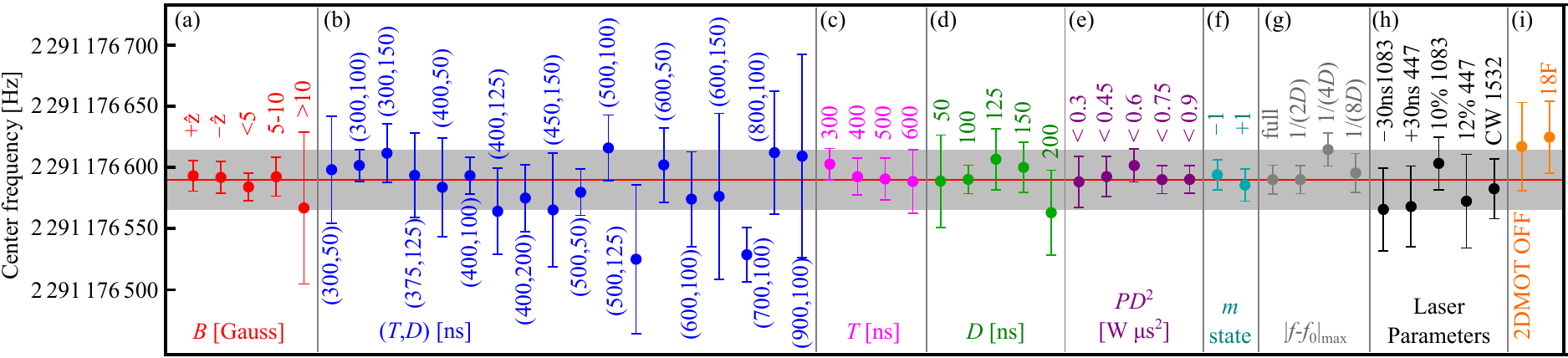}
\caption{\label{fig:Parameters} 
A summary of the average obtained center
value for the various values of the experimental parameters. 
Details are explained in the text. 
All points use the saturation restriction of
$PD^2$$<$0.75~W$\mu$s$^2$, the 
frequency range 
$|f-f_0|$$<$$1/(2D)$, 
and the magnetic field restriction
$|\vec{B}|$$<$10~gauss,
unless otherwise specified.
}
\end{figure*}

Our experiment is performed within
a magnetic field 
$\vec{B}$
of 
typically 
5~gauss. 
This
$\vec{B}$
is applied by 
20-cm-radius
Helmholtz coils,
with 
geomagnetic and other local fields
canceled by six larger coils.
The largest systematic effect
in our measurement is a 
second-order
Zeeman 
shift
of
429.5~Hz/gauss$^2$.
The quadratic shift rate is 
precisely calculated
\cite{yan1994high},
and has been directly
tested by other measurements
\cite{lewis1970experiments,
PRL.95.203001}
using larger 
$B$. 
We also use
larger 
$B$
to directly show that
we understand the magnetic shifts
at a level of 
$<$0.1\%,
and we include a 
0.1\% uncertainty 
to all 
Zeeman 
corrections.
Fig~\ref{fig:Parameters}(a)
shows that that measurements
taken with
$\vec{B}$ in the 
$+\hat{z}$
and
$-\hat{z}$
directions agree,
and that those with  
$|\vec{B}|$$<$5~gauss
agree with those taken for 
$|\vec{B}|$$>$10~gauss
(which have, 
on average,
a 
six times 
larger
Zeeman 
shift).

Eq.~(\ref{eq:LineShape}) assumes 
perfect microwave pulses, 
including sudden 
turn-on
and 
turn-off,
no chirp in the 
phase due to the 
microwave switching,
and no changes in intensity
or phase profiles as 
a function of 
$f$.
Imperfections in the pulses
cause the second largest
systematic in our measurement.
In our previous SOF measurement
\cite{PRA.79.060503},
we monitored the microwave 
pulses by directly digitizing them with an
oscilloscope and attempted to determine
shifts that result from pulse imperfections
by integrating the
Schr$\ddot{\rm o}$dinger
equation using the recorded microwave pulses.
Because of the excellent
S/N,
in this work,
we could test such corrections directly, 
by varying both the power
$P$
and the magnitude of the imperfections.
Unfortunately, 
for both 
SOF
and 
FOSOF
measurements,
we find the calculated corrections to be 
unreliable, 
with the corrected centers from 
different distortions
and 
$P$ 
disagreeing 
at the level of several hundred 
Hz.
We attribute this inconsistency to 
a combination of two factors:
that the oscilloscope does not faithfully
render the microwave pulse sent to it,
and that the atoms do not see the same
microwave profile as the oscilloscope
(due, 
e.g., 
to impedance mismatches and 
resulting reflections and interferences).

\begin{figure}[t!]
\includegraphics[width=3.3in]
{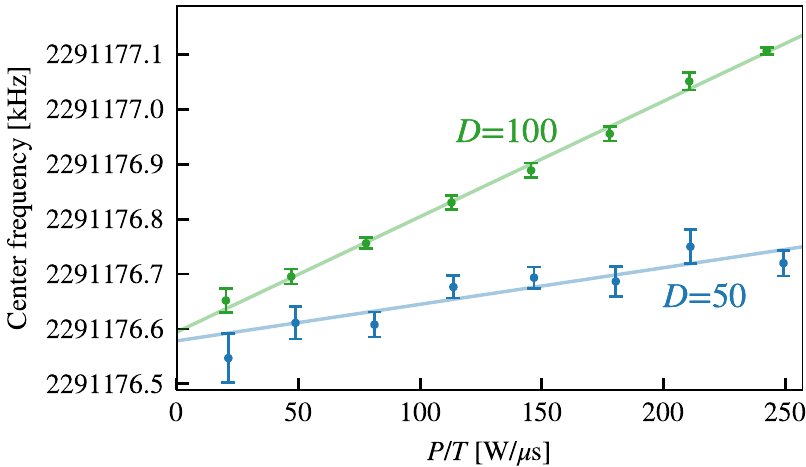}
\caption{\label{fig:LinearShifts} 
The extrapolation of the averaged
$D$$=$100~ns
and 
$D$$=$50~ns
FOSOF fit centers to 
$P$$=$0,
where the center is unaffected by 
imperfections in the pulses.
}
\end{figure}

Extensive modeling,
however,
shows that 
any form of distortion 
gives shifts that 
vary linearly with
$P$.
We find that 
FOSOF 
shifts extrapolate
exactly to zero in the  
TDPT 
($P$$\to$0)
limit,
but
SOF 
shifts 
are only within
$\sim$200~Hz
of zero
for 
$P$$\to$0.
As a result of this modeling, 
the strategy used here 
is to extrapolate our 
FOSOF centers to 
$P$$=$0,
as shown,
e.g., 
in 
Fig~\ref{fig:LinearShifts}.
These extrapolations also 
account for small 
($<$40~Hz)
AC
Zeeman
shifts,
and for
AC
Stark shifts,
which are even smaller here
because the beam passes
through the standing wave
at a node for the microwave 
electric field 
(an 
anti-node 
for the magnetic field).
Measurements are repeated  
for combinations of
$T$
and
$D$
to confirm that all sets of parameters 
extrapolate to a single 
intercept,
as shown in 
Fig.~\ref{fig:Parameters}(b).
Fig.~\ref{fig:Parameters}(c)
and 
(d)
give averages for each
$T$
and
$D$,
respectively. 
Both in the modeling and data,
the extrapolations have slopes that 
vary approximately as
$D/T$,
and we could fit the data by using 
one extrapolation constant for each
$D$ 
used in this experiment
(as shown in  
Fig~\ref{fig:LinearShifts}).
Some data are also taken with
larger imperfections 
which led to 
three-times-larger 
slopes for all extrapolations, 
but still obtained consistent intercepts
within the 
100~Hz accuracy of the test.
As 
$P$
increases, 
the 
FOSOF 
signal starts to saturate, 
and the shifts no longer follow
a linear trend.
We use only data 
well below saturation in our fits,
Fig.~\ref{fig:Parameters}(e)
shows that our results
are independent of how strongly
we enforce this saturation limit
(by restricting 
$P\,D^2$$<$0.9, 
0.75,
0.6,
0.45,
or
0.3~W$\mu$s$^2$).

It is clear that our previous 
SOF
data 
\cite{PRA.79.060503}
should not have been
corrected based on the oscilloscope
traces, 
but rather should have been extrapolated
to 
$P$=0.
Using this method, 
the result from that work
\cite{PRA.79.060503}
changes from 
$2\,291\,177\,530(350)$~Hz
to
$2\,291\,176\,655(660)$~Hz.

The polarization for the 
optical-pumping 
step is reversed for half of
the measurements, 
allowing the 
$2^3$P$_1(m$=$+1)\!\!\to\!2^3$P$_2(m$=$+$1)
transition to also be measured.
Fig.~\ref{fig:Parameters}(f)
shows that consistent results are 
obtained with 
$m$$=$$\pm1$.
To test for possible 
FOSOF lineshape effects, 
the data are also refit with only the 
central frequencies  
($|f$$-$$f_0|$$<$$1/(2D)$,
$1/(4D)$,
or
$1/(8D)$)
included in the fit.
Consistent results are found, 
as shown in 
Fig.~\ref{fig:Parameters}(g).
Data taken with up to 
25 
times higher pressure
show no indication of a pressure
shift and limits a possible shift to 
$<$4~Hz.

One concern is that the 
18$^3$P 
atoms
travel through the rest of the 
airline before being ionized,
and therefore they 
are exposed to additional 
microwave pulses which
could drive  
atomic processes involving these
states. 
To test for a systematic shift due 
to such processes, 
measurements are taken 
with lower duty cycles, 
with sufficient time between the 
FOSOF cycles 
to allow the 
18$^3$P
atoms to exit without seeing 
additional microwave pulses.
Additionally,
measurements are performed in which the 
laser and microwave excitations are
moved to the other side of the inner conductor 
of the
airline of 
Fig.~\ref{fig:ExptSetup}(c),
so that the 
18$^3$P
atoms spend far more time in the 
microwave fields.
Finally, 
measurements are performed using 
the 
18$^3$F
state
instead of
18$^3$P
state
(Fig.~\ref{fig:Parameters}(h)).
All three tests show that the
$n$=18 
states play no significant role.

To test for light shifts due to unintended 
temporal overlap of the laser and microwave
pulses, 
data are taken at 
8 times
smaller 
447-nm 
power,
10 times
smaller
1083-nm 
power,
with the 
1532-nm 
laser on throughout the 
whole measurement sequence 
of 
Fig.~\ref{fig:ExptSetup}(d),
and by taking 
data with larger time delays between
the laser and microwave pulses.
In all cases,
Fig.~\ref{fig:Parameters}(i),
the results indicate no light shifts.
Consistent centers 
are also found
for different offset frequencies:
-2.8~kHz$<$$\delta\!f$$<$2.8~kHz.
Also, 
different source temperatures
(110~K 
to
300~K),
different currents driving the 
DC 
discharge 
(10~mA
to 
25~mA), 
and not using the 
2DMOT 
(Fig.~\ref{fig:Parameters}(h))
reveal no inconsistencies.

Since this measurement is a first
demonstration of the 
FOSOF
method,
we performed a parallel 
SOF experiment, 
in which we used the same 
microwave system as the one we 
used in 
Ref.~\cite{PRA.79.060503}, 
and the same laser cooling and
detection method as applied here.
Also, 
the magnitude and phase of the 
sinusoidal signals seen in the present
FOSOF 
measurements 
(e.g., those in 
Fig.~\ref{fig:LineShape}(b) and (c))
can be used to construct 
SOF data points,
and these points can be fit to an
SOF
lineshape to find 
$f_0$.
The $f_0$ obtained in this manner
are less precise than
the 
FOSOF 
$f_0$.
Results for both
SOF
analyses 
(when extrapolated to 
$P$$=$0)
agree with our 
present result to within 
200~Hz -- 
the level of agreement that
we would expect since our modeling 
shows that 
SOF 
centers have a residual
systematic shift even at 
$P$$=$0.

\begin{figure}[t!]
\includegraphics[width=3.5in]
{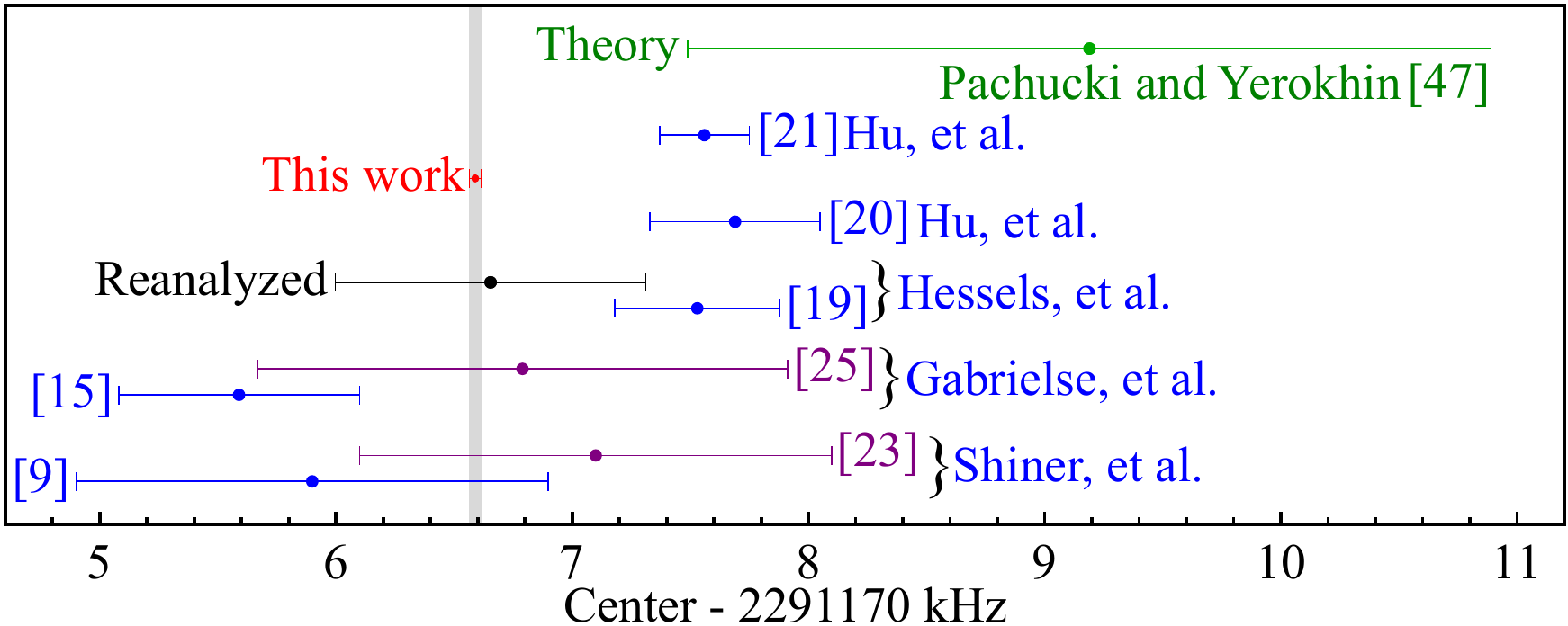}
\caption{\label{fig:Compare} 
A comparison of the present measurement to
previous measurements 
\cite{zheng2017laser,
PRA.91.030502,
PRA.79.060503,
PRL.95.203001,
PRL.105.123001}
and to theory
\cite{PRL.104.070403}
Corrected centers including 
quantum interference effects
\cite{PRA.86.040501,
marsman2015quantum}
are also shown.
The reanalyzed center from our previous 
SOF 
measurement 
as discussed in this work,
is also shown.
}
\end{figure}

The weighted average of the 
results shown in 
Fig.~\ref{fig:Parameters}(b)
is
$2\,291\,176\,590(11)_{\rm ext}(8)_{B}$~Hz,
where the two uncertainties come from the 
extrapolations to 
$P$$=$0 
and from 
the 
Zeeman
shift. 
Based on the level of 
consistency demonstrated for 
a wide range of parameters in 
Fig.~\ref{fig:Parameters},
we conservatively assign a
larger uncertainty of 
$\pm$25~Hz 
to our measurement,
giving a final measurement result
of
\begin{equation}
[E(2^3{\rm P}_1)-E(2^3{\rm P}_2)]/h
=2\,291\,176\,590(25)~{\rm Hz}.
\end{equation}
Our result is   
slightly smaller
(1.5~times 
the estimated theoretical uncertainty)
than the best theoretical prediction
\cite{PRL.104.070403},
as seen in 
Fig.~\ref{fig:Compare}.
It disagrees with recent laser 
measurements 
by  
Hu, et al.
\cite{zheng2017laser,PRA.91.030502}
by 
4.9
and 
2.9 
times their uncertainties.
Only after the correction applied in this 
work does our previous
SOF
measurement
\cite{PRA.79.060503}
agree with the present measurement.
With the inclusion of quantum interference
corrections  
\cite{marsman2015quantum,
PRA.86.040501},
the 
saturated-absorption 
measurement of 
Gabrielse, et al.
\cite{PRL.95.203001}
and the the 
laser measurement of 
Shiner, et al.
\cite{PRL.105.123001}
also agree with the present measurement.

This measurement is the most precise
measurement to date of helium fine structure,
and represents a major advance in this precision.
The outstanding 
signal-to-noise
ratio has allowed for a very extensive survey of 
systematic effects.
This work sets the stage for the a new level of 
accuracy for this fine structure, 
which, 
when combined with more precise theory,
could provide 
$<$1~ppb
tests of the physics and constants 
relevant to the interval --
including a precise determination of the 
fine-structure constant,
the most precise test of 
QED 
in a multi-electron system,
and tests for physics beyond the 
Standard Model.

This work is supported by 
NSERC, 
CRC, 
ORF, 
CFI, 
NIST 
and a 
York University Research Chair. 
We thank 
Amar Vutha, 
Daniel Fitzakerley, 
Matthew George, 
Hermina Beica,
Travis Valdez,
Nikita Bezginov,
and 
Cody Storry 
for their contributions to this work.

\newpage
\bibliography{He2018SmallInterval}

\end{document}